# ON THE FIRST ANTHROPIC ARGUMENT IN ASTROBIOLOGY

Milan M. Ćirković

*Astronomical Observatory Belgrade*

*Volgina 7, 11160 Belgrade, Serbia and Montenegro*

*e-mail:* mcirkovic@aob.aob.bg.ac.yu

**Abstract.** We consider the little-known anthropic argument of Fontenelle dealing with the nature of cometary orbits, given a year before the publication of Newton's *Principia*. This is particularly interesting in view of the rapid development of the recently resurgent theories of cometary catastrophism and their role in the modern astrobiological debates, for instance in the "rare Earth" hypothesis of Ward and Brownlee.

**Keywords:** comets:general—stars: planetary systems—extraterrestrial intelligence—history and philosophy of astronomy

## 1. Introduction

The issue of stability of conditions prevailing on (at least potentially) habitable planets throughout the Galaxy is the central question of the nascent science of astrobiology (e.g. Darling 2001). We are lucky enough to live in an epoch of great astronomical discoveries, the most distinguished probably being the discovery of dozens of planets orbiting nearby stars (for nice reviews, see Marcy and Butler 1998, 2000; Ksanfomaliti 2000). This particular discovery brings about a profound change in our thinking about the universe, and prompts further questions on the frequency of Earth-like habitats elsewhere in the Galaxy. In a sense, it answers a question posed since antiquity: are there other, potentially inhabited or inhabitable, worlds in the vastness of space? In asking that question, obviously, we take into account our properties as intelligent observers, as well as physical, chemical and other pre-conditions necessary for our existence. The latter are the topic of the so-called anthropic principle(s), the subject of much debate and controversy in cosmology, fundamental physics, and philosophy of science.

Arguably the most significant construal of the anthropic principle(s) is the one having to do with restrictions that the existence of human beings places on the observable features of the universe (Carter 1974, 1983; Barrow and Tipler 1986; Bostrom 2002). In other words, the anthropic principle acts as an *observation selection effect* in the set of all possible observations of the universe. However, the very definition of the universe evolved, of course (cf. Munitz 1986). The universe of the standard ("Big Bang") cosmological model is drastically different from the essentially static and eternal (or supernaturally created in essentially the same state) universe of previous epochs, notably the one of Bruno, Descartes, Halley, Newton, Laplace, Lord Kelvin, and early Einstein. In attempting to reconstruct the evolution of cosmological and cosmogonical thinking, one should be very careful to avoid conflating various levels of cosmological discourse with which we are familiar today, with those known in earlier times. On the contrary, a conclusion of some generality arrived at in the times of much poorer level of understanding and description is certainly to be exceptionally praised from both the historical and the epistemological points of view.

In this note, it is our goal to show that one of the very first instances of anthropic thinking, applied to planetary science and astrobiology (in modern terms), occurred very early, in the work of French philosopher and naturalist Bernard Le Bouyier de Fontenelle (1657-1757), published in 1686, a single year before the great scientific revolution inaugurated by Newton's *Principia*. Fontenelle (Figure 1) was one of the most important precursors of the Enlightenment. In his long and productive career, he touched upon practically all aspects of human knowledge, other books of his bearing titles such as *On the Usefulness of Mathematical Learning* or *A History of Oracles*. Of course, the recent Copernican revolution in astronomy and cosmology could not escape his attention, and he supported the new paradigm with more than his usual eloquence and wit in *Conversation on the Plurality of the Worlds*, where the argument presently under scrutiny is located.

Fontenelle's argument deals with the comets, bodies which presented an incredibly popular scientific topic in his day, and which have remained in the focus of interest of planetary studies until now. It was already known, through the great studies of Halley and others (see Yeomans 1991; Schechen 1997) that comets travel around the Sun on very elongated orbits which have high inclinations, that is, their orbital planes are at large angles from the orbital planet of Earth (and other planets known at the time). The late XVII century was in many ways obsessed with these strange celestial visitors and with good reason (see, for instance, a colorful "catalogue" of bright comets in Figure 2). For important epistemological and cultural implications of comet observations in the relevant period, see an excellent study by Barker and Goldstein (1988). The unusual orbital behavior of comets, in eyes of Fontenelle and his contemporaries (as well as of modern cosmogonists!) is strange and requires an explanation. He offered an unusual, ingenious, and even today controversial argument.

## 2. Fontenelle's argument

The remarkable argument put forward by Fontenelle in 1686, a single year before the appearance of Newton's majestic *Principia* and ten years before Whiston's *A New Theory of Earth*,[1] is essentially contained in a single paragraph of his *Conversation on the Plurality of the Worlds*. It reads (Fontenelle 1767):

> In the next places, the reason why the planes of their [comets'] motions are not in the plane of the ecliptic, or any of the planetary orbits, is extremely evident; for had this been the case, it would have been impossible for the Earth to be out of the way of the comets' tails. Nay, the possibility of an immediate encounter or shock of the body, of a comet would have been too frequent; and considering how great is the velocity of a comet at such a time, the collision of two such bodies must necessarily be destructive of each other; nor perhaps could the inhabitants of planets long survive frequent immersions in the tails of comets, as they would be liable to in such a situation. Not to mention anything of the irregularities and confusion that must happen in the motion of planets and comets, if their orbits were all disposed in the same plane.

---

[1] The book which is usually celebrated as the first serious attempt at formulating a theory of cometary catastrophism.

Thus, to the question: *why are (observed) orbits of comets highly inclined, in contradistinction to the coplanar planetary orbits?* Fontenelle offers a deceptively simple answer. We would not be here – to contemplate on the peculiarities of cometary trajectories – if these orbits were different (that is, similar to those of planets). This had been published 8 years before celebrated Halley's suggestion of December 12, 1694, that comets might collide with planets (Halley 1726):

> This is spoken to Astronomers: But, what might be the Consequences of so near an Appulse; or of a Contact; or, lastly, of a Shock of the Coelestial Bodies, (which is by no means impossible to come to pass) I leave to be discuss'd by the Studious of Physical Matters.

This famous idea has been followed up by such luminaries as Newton, Wright, Laplace, Lagrange, and others, in the vein of what is usually (and only partially justifiable) called "Biblical catastrophism" of the pre-Cuvier epoch.[2] Fontenelle wrote the passage more than 18 years before Newton wondered (in *Opticks*):

> Whence is it that planets move all one and the same way in orbs concentrick, while comets move all manner of ways in orbs very excentric... blind Fate could never make all the planets move one and the same way in orbs concentrick, some inconsiderable irregularities excepted, which may have risen from the mutual actions of comets and planets upon one another, and which will be apt to increase, till this system wants a reformation. Such a wonderful uniformity in the planetary system must be allowed the effect of choice.

Thus Newton, as a great promoter of the Design argument in natural philosophy, failed to understand the power of the Fontenelle's argument, and went deeper into a blind alley (from the modern point of view) of seeking the supranatural Design and/or regulating mechanism. The same tension between the apparent design and the explanatory

---

[2] For colourful pieces of its history, see Clube and Napier 1990.

"filtering" through various observation selection effects persists to this day, and is the source of innumerable debates and confusions. We shall return to this point below.

Unfortunately, in the subsequent long dogmatic slumber of uniformitarian domination in the entire realm of natural sciences, it has been too often forgotten what is and what is not "by no means impossible to come to pass". In the present era of revived "neocatastrophism" (e.g. Schindewolf 1962; Clube and Napier 1982, 1984, 1990; Clube 1995; Raup 1999), some of these early thinkers (like Halley) received a renewed attention, but Fontenelle is undeservedly rarely mentioned. It is one of the purposes of this note that this historical injustice is at least partially rectified.

Now, there are two arguments in the quoted passage of Fontenelle. The first concerns the dynamical influence through impacts. Impacts of comets upon planets would have been much more frequent and destructive if their orbits lay in the plain of the ecliptic. Such collisions would be highly destructive to lifeforms (especially advanced and intelligent ones). As we shall see below, its validity has been fortified by scientific findings in recent decades. The second argument concerns the effects of immersion of inhabited planets in cometary tails. The standard modern view is that this does not make sense, since such phenomena are thought to be completely harmless, due to the excessively low density of cometary tails. Still, there are some dissenters from this view, the most famous being the late Sir Fred Hoyle, who (with Chandra Wickramasinghe) argued, rather notoriously, that comets may be vehicles for panspermia, and even cause familiar cases of epidemic diseases (Hoyle and Wickramasinghe 1979; Hoyle, Wickramasinghe, and Watkins 1986). Another suggested effect is a possible climatic influence due to depositing of cometary dust, with the zodiacal cloud as an intermediate reservoir, in the Earth's atmosphere (Hoyle 1987; Napier 2001). It would be only prudent to state that the effects of prolonged interaction of a cometary tail and the Earth have not been meticulously studied so far. In any case, there is obviously *less interaction* between Earth and comets due to this specific feature (high inclination) of the latters' orbits in comparison to the counterfactual case Fontenelle considers.

Now, what could be counted as an *explanation* of this particular feature? When we come to the "grand questions" on the origin of the universe (or "world," construed in some narrower sense!), we are left with surprisingly few viable options. One of them, certainly unsatisfactory, is to claim some supernatural or Divine cause which is not accessible to further elucidation. Another, and it has become more and more interesting as modern cosmology progressed, is that any atypical particular feature can be made unsurprising and "natural" by embedding it in a set or distribution broad enough to include many (or all) cases of the phenomenon in question. This is the usual approach of the anthropic thinking. Thus, Fontenelle may appeal to a kind of "principle of fecundity" (Nozick 1981): we explain the observed feature of a system by embedding its specific features into a wider system where many (or all) possibilities are realized. As Nozick (1981, p. 131) writes:

> The principle of fecundity is an invariance principle. Within general relativity, scientific laws are invariant with respect to all differentiable coordinate transformations. The principle of fecundity's description of the structure of possibilities is invariant across all possible worlds. There is no one specially privileged or preferred possibility, including the one we call actual... The actual world has no specially privileged status, it merely is the world where we are. Other independently realized possibilities also are correctly referred to by their inhabitants as actual.

Of course, Nozick speaks of it in the language of modern metaphysics, but we need to remember that the ontological construal of the locution "world" has undergone revolutionary changes since the epoch of Enlightenment. The universe of Fontenelle was essentially the universe of Bruno: an ensemble of different *planetary systems* taken as the self-contained ontological units. That Fontenelle's worlds were planets (solar and extrasolar) and that his main interest was astrobiological is testified by the following words from his Preface:

> I have chosen that part of Philosophy which is most like to excite curiosity; for what can more concern us, than to know how this world which we inhabit, is made; and whether there be any other worlds like it, which are also inhabited as this is?

(Parenthetically, these words have been used as motto by one of the earliest astrobiological treatises in XX century, "Life on Other Worlds", by Sir Harold Spencer Jones, the Astronomer Royal; see Spencer Jones 1952). As we today—inspired by inflation and other developments in quantum cosmology—speak of topologically disconnected cosmological domains ("universes") within a larger whole (the "multiverse"; cf. Linde 1990), thus Fontenelle spoke on the differently arranged *planetary* domains.

And the multitude of worlds (however rationally construed) gives an excellent physical basis for application of the anthropic selection effect. Among many worlds, the distinction between inhabited and uninhabited is readily made. And the issue of perception of peculiarity of inhabited subset in the entire set is a legitimate target of physical inquiry. In other words, we have three possible options of explaining the peculiar (non-planar) nature of cometary orbits in *our* planetary system. The first is to deny the validity and meaningfulness of the question; this is the standard theistic answer which forbids further discussion. Beside the unacceptable epistemological nature of this answer, one should mention that it is today completely irrational to apply it to such small subsystems as the Solar system, since we know that they form as part of a much larger whole. As to the origin and properties of this larger whole (the Galaxy) we do have different (and working!) explanations. From the other two options, one—causal— entails the idea that there is a law-like reason (presumably to be derived from the future "Theory of Everything" or some other high-level physical theory) for atypical or surprising structure of the early Solar system. In other words, an enormous amount of information necessary for description of the atypical initial conditions can be encoded in some new law(s) of nature and consequent law-like correlations of various matter and vacuum fields. This option is still viable for *cosmology*, but hardly for planetary *cosmogony*. Cosmogonical initial conditions are not privileged in any way over initial conditions for any other physical process; we do not seek an explanation for (say) the formation of chemical elements in a future unified field theory. We seek it much lower down on the epistemological ladder. The other—anthropic—option is to avoid giving a specific description through embedding those conditions into a sufficiently symmetric ("typical") background. Again, stated in terms of information, the same long description of what we perceive as atypical initial conditions arises—as so often in physics!—from the process of *symmetry breaking*. The overall description is simple enough, and may be reduced (in the extreme case) to a rule similar to "All possible combinations of initial conditions exist." That such a high degree of symmetry can indeed completely reproduce the situation in our particular cosmological domain becomes an immediate consequence (cf. Tegmark 1996; Collier 1996).

Let us now think of the application of this mode of thinking to the particular cosmogonical issue. We would similarly say "planetary systems with all possible configurations of planetary and cometary orbits exist" as a part of the larger whole (say our Galaxy). Now, we ask: are all such configurations compatible with our existence (on a planet!)? And the answer, intuitively clear even to Fontenelle, with his rudimentary understanding of preconditions for complex life and intelligence, seems to be negative. There is only a subset of all configurations leading to the emergence of us as intelligent observers, namely the one in which collisions between planets and smaller bodies (comets and asteroids) are not too frequent. Thus, our existence acts as an observational selection effect (cf. Bostrom 2002), or "filter" selecting those sites – in this case planetary systems – where configurations of cometary orbits are in some sense atypical.

It is interesting to speculate whether the rather low regard in which many philosophers of science hold anthropic arguments originates in the fact that from about the time of Newton till a couple decades ago, the dominating view emerging from natural sciences presumed a single world, in both planetary and cosmological connotations. The history of ideas might have been very different, according to this line of thought, if the fecundity principle had gone unchallenged from Bruno's days until the present, when it was resurrected on a truly universal scale in the modern theories of chaotic inflation of Linde, Vilenkin, and others.

**3. Cometary catastrophism resurgent?**

As of May 25, 2003, there are 108 extrasolar planets in 94 planetary systems detected (for regularly updated list, see http://www.obspm.fr/planets). We remain in virtually complete ignorance about the details of those planetary systems, notably their stability and their local cometary environments. What we are becoming aware of is the fact that simplified cosmogonic theories about the planetary systems' origination do not work; among the newly discovered planets there are quite a few oddities, like Jovian planets at a fraction of an AU from the central star, or planets with large orbital eccentricities. Therefore, we may toy with various thought experiments; suppose, for instance, that each planetary system has its own comets and that the distribution of average inclination of cometary orbits is a random one. We expect the frequency of cometary collisions with planets to be inversely proportional to the maximal inclination of the set of comets (thus being maximal in the case of coplanar orbits, as Fontenelle envisaged).

In the recently very widely discussed hypothesis of "rare Earth" (Ward and Brownlee 2000), the role of cometary and/or asteroidal bombardment is essential. Namely, Ward and Brownlee argue that, while simple lifeforms may be ubiquitous throughout the Galaxy, complex life (and especially intelligent life) must be very rare, not only since we do not see any trace of it (Fermi's "paradox"; Brin 1983), but also because we now have sufficient knowledge to assess the various "fine tunings" necessary for emergence of astrobiological complexity (chemical evolution, orbital stability, climate stability, plate tectonics, etc.).

One of the main factors in keeping an equilibrium state between growth of complexity of life forms on the surfaces of terrestrial planets and their destruction or prevention of growth is the phenomenon of cometary/asteroidal bombardment of planetary surfaces. And today we know that cometary/asteroidal bombardment is governed by the dynamics of Oort clouds of individual planetary systems. Obviously, if the impact rate is above a certain threshold level, complex life has no chance of arising ("impact frustration"; Schopf 1999; Ward and Brownlee 2000). This idea has been corroborated by the almost consensual view that some of the major biological mass extinctions, notably the KT mass extinction 65 Myr ago, have been caused by asteroidal/cometary impact events (e.g. Raup 1986, 1999). In addition, it is widely recognized that complex life is more susceptible to extinction than early, simple lifeforms.

A sizeable literature has been devoted so far to the potentially lethal processes triggered by a major cometary impact. Some reviews of thought of "the Studious of Physical Matters" can be found in the bibliography. In addition, some thoughts on the impact of this "neocatastrophic" worldview on wider cultural issues have also been published (e.g. Bailey 1995; Clube 1995). One of the most interesting has been published under the indicative title "The Fundamental Role of Giant Comets in Earth's History" (Clube 1992). In any case, it seems that after the long uniformitarian slumber, we are gradually returning to an awareness of the significance of interactions between us and our "cometary environment" (Clube and Napier 1984; Napier and Clube 1997).

In these circumstances, Fontenelle's anthropic argument gains a new significance. If the main reason why complex life is rare in the Galaxy is indeed a high average sensitivity of planetary biospheres to cometary bombardment, the fact that the impact rates during the last Gyr of the terrestrial history were so low as to enable the rise of intelligence is really even more atypical than could have been imagined in the time of Newton. A prediction of the "rare Earth" hypothesis is that, on the average, impact rates in other planetary systems (for instance, the newly detected ones) are much higher, and the Solar System is indeed an exception important – for its biological consequences – on the Galactic level. Instead of a crude Design argument, we might, following Fontenelle, employ the observation selection and thus give a rational (albeit pessimistic) reply to the perennial SETI question: Where are they? (cf. Hart 1975; Carter 1983)

## 4. Anthropic selection vs. design

There are a lot of ongoing debates on the interpretation of anthropic coincidences or "fine-tunings," but the debates are usually plagued by more than a few misunderstandings and confusing issues. Some of the misunderstandings and confusions stem from misunderstanding of the nature of the explanatory task itself. As an instance of the latter we can consider the effort invested in proving that the anthropic "coincidences" or "fine-tunings" do not imply (intelligent) design.[3] As explained in numerous beautiful details and examples in the recent monograph of Bostrom (2002), the true spirit of anthropic reasoning has nothing to do with teleological or the theistic design agendas. Instead, it is dealing with exactly the sort of observational selection effects discussed above; and its modern explication is surprisingly similar to the one offered by Fontenelle more than three centuries ago (albeit on a much smaller level).

It is interesting that Fontenelle's writing is quite bold in the emphasis on the explanatory efficiency of the anthropic selection effect: "the reason... is extremely evident." In contradistinction to the host of modern authors, who only reluctantly, and with great hesitation admit this anthropic point of view as valid, Fontenelle is refreshingly clear and reasonable. The cause is perhaps the fact that he was living and working in the epoch of great changes and strides forward in the human understanding of the universe, to which a dialogue between science and metaphysics was quite natural and indeed a necessary part of any discourse in natural philosophy. In particular, he lacked a bizarre fear of anthropocentrism, which permeates modern writings on the anthropic principles. In particular the opponents of the anthropic reasoning have raised accusations along that favorite line of attack (e.g. Pagels 1998). These accusations are not only *quid pro quo*, but also based essentially on innuendo: serious anthropic thinking has nothing to do with anthropocentrism, except the unfortunate similarity of words. Fontenelle's argument demonstrates this quite clearly.

---
[3] An egregious recent example is Klee (2002).

It was Newton's cosmology, ironically enough, which was narrow and anthropocentric, and with good reason, since one of the main strands of Newton's work was his aggressive theistic agenda and promotion of the Design argument. The same tension between Newton's and Fontenelle's views is still present today in the form of the tension between the two interpretations of the anthropic fine-tunings. The interpretation of these "coincidences" as indications of intelligent design represents a continuation of the broadly Newtonian worldview, while the antithetical view of fine-tunings as observation selection effects (within the multiverse of either cosmology or quantum mechanics) bears the hallmark of Fontenelle's approach. The former is still motivated by some extra-scientific reasons as it was in the XVII century, while the latter again displays scientific superiority.

## 5. Conclusions

We have investigated the first instance of application of the observer self-selection in what would centuries later become astrobiology, notably the issue of cometary bombardment and impact frustration of the origin and evolution of life and sentience on Earth. Fontenelle's argument for a high inclination of cometary orbits is an excellent illustration of the entire paradigm of regarding the *anthropic principle as selection* *effect*, having nothing to do with the alleged metaphysical notion of intelligent design. Thus, its value is not only historical, but pedagogical as well.

**Acknowledgements.** The author wholeheartedly thanks Branislav Nikolić, Saša Nedeljković, Milan Bogosavljević, Ivana Dragićević, and a referee for invaluable help in improving a previous version of the manuscript. The inspiration and kind support of Irena Diklić have been crucial for the completion of this project. The author also acknowledges partial support of the Ministry of Science, Technology, and Development of Serbia through the projects no. 1196, "Astrophysical Spectroscopy of Extragalactic Objects" and no. 1468, "Structure and Kinematics of the Milky Way."

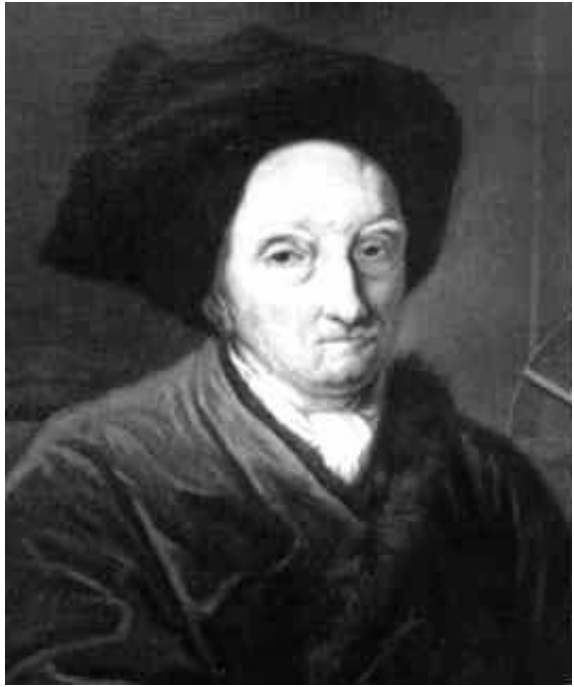

**Figure 1.** Bernard Le Bouyier de Fontenelle (1657–1757).

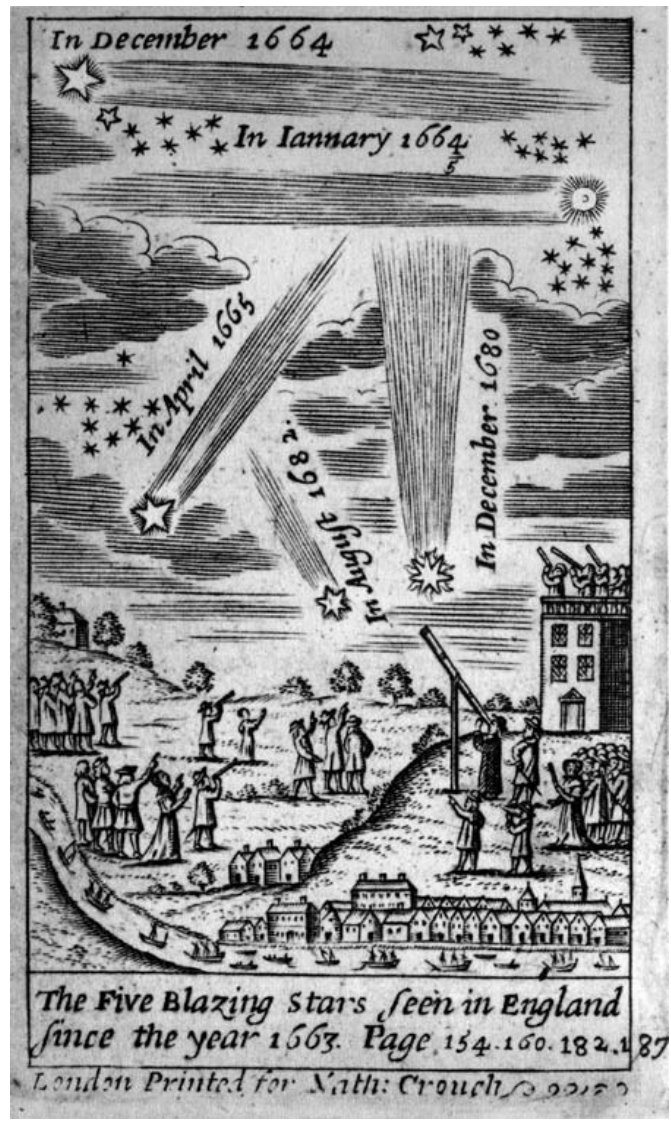

**Figure 2.** The obsession of late XVII century with comets is summarized in this vivid illustration from the book (or pamphlet) on "Surprizing Miracles" by Nathaniel Crouch (London, 1685).[4]

---

[4] Courtesy of Houghton Library, Harvard University.